\documentclass[prb,twocolumn,superscriptaddress,showpacs,showkeys]{revtex4}% Physical Review B

\usepackage{graphicx} 
\usepackage{dcolumn} 
\usepackage{bm}

\begin{document}

\preprint{APS/123-QED}

\title{Neutral-ionic phase transition : a thorough ab-initio 
study of TTF-CA\\}

\author{V. Oison}
\author{C. Katan}%
 \email{Claudine.Katan@univ-rennes1.fr}
\author{P. Rabiller}
\affiliation{%
Groupe Mati\`ere Condens\'ee et Mat\'eriaux,UMR6626 CNRS - 
Universit\'e Rennes 1,\\
Campus de Beaulieu B\^at. 11A, F-35042 Rennes Cedex, France
}%
\author{M. Souhassou}
\affiliation{LCM$^3$B, UMR7036 CNRS - Universit\'e Henri 
Poincar\'e Nancy 1, F-54506 Vand\oe{}uvre l\`es Nancy, France}
\author{C. Koenig}
\affiliation{%
Groupe Mati\`ere Condens\'ee et Mat\'eriaux,UMR6626 CNRS -
Universit\'e Rennes 1,\\
Campus de Beaulieu B\^at. 11A, F-35042 Rennes Cedex, France
}%

\date{\today}% It is always \today, today,

\begin{abstract}

The prototype compound for the \textit{neutral-ionic} phase
transition, namely TTF-CA, is theoretically investigated by
first-principles density functional theory calculations. The
study is based on three neutron diffraction structures collected 
at 40, 90 and 300~K (Le~Cointe et al., Phys. Rev. B \textbf{51}, 3374
(1995)). By means of a topological analysis of the total charge
densities, we provide a very precise picture of intra and inter-chain 
interactions. Moreover, our calculations reveal that the thermal
lattice contraction reduces the indirect band gap of this organic
semi-conductor in the neutral phase, and nearly closes it in the 
vicinity of the transition temperature. A possible mechanism of the 
\textit{neutral-ionic} phase transition is discussed. The
charge transfer from TTF to CA is also derived by using three
different techniques.

\end{abstract}


\maketitle

\section{Introduction:}

Charge transfer salts presenting mixed stacks with alternating
donor (D) and acceptor (A) molecules have been extensively studied 
over the last 20 years for their original neutral-ionic phase
transitions (NIT).\cite{mayerle, torrance}
Very recently, this class of materials has gained 
renewed interest as it has been demonstrated experimentally that
in some cases
the conversion from the neutral state (N) to the ionic state 
(I) or from I to
N can also be induced by photo-irradiation.\cite{koshihara}
Despite intensive
theoretical work, the mechanism of the phase transition and 
photo-conversion has not yet been clarified. In these systems, 
unlike in
other classes of compounds, no dominant interaction has been evidenced
and the nature of the NIT must be related to a subtle interplay of 
different type of interactions.

Depending on the choice of D and A molecules as well as on
molecular packing, different type of NIT have been observed mainly under
pressure and sometimes also under temperature variation. Continuous 
and discontinuous NIT have been reported 
presenting different amplitudes of charge
transfer (CT) variations from D to A,
often with a \textit{dimerization} in DA pairs along the stacking axis.
The TTF-CA complex made from tetrathiafulvalene (D = TTF)
and p-chloranil (A = CA) molecules is considered as the prototype
compound for NIT. At atmospheric pressure, it undergoes a first order
NIT at a critical temperature of about 80~K. 
This symmetry
breaking phase transition leads to a ferroelectric low temperature phase
where the initially planar D and A molecules are both deformed and displaced
to form DA pairs along the stacking chains.\cite{lecointe} 
According to vibrational spectroscopy~\cite{girlando2} 
and CT absorption spectra~\cite{jacobsen} the CT has been estimated 
to be about 0.2 e$^-$ in the neutral high temperature
phase and 0.7 e$^-$ in the ionic low temperature phase. 
TTF-CA is also a typical example showing N-I and I-N conversion
under photo-irradiation.\cite{koshihara} Recently, ultrafast (ps) 
optical switching from I to N has been observed by femtosecond 
reflection spectroscopy.\cite{iwai}

Numerous theoretical approaches have been developed to study the NIT. 
For many years, the balance between the energy needed to ionize molecules 
and the gain in the Madelung energy has been at the heart of the concept 
of NIT.\cite{torrance,iwai} 
More advanced models mainly concern extensions of the one-dimensional (1D)
modified Hubbard model which
includes on-site Coulomb repulsion, transfer integral and on-site energy.
Anusooya-Pati et al.\cite{anusooya} have recently shown 
that such a model has a 
continuous NIT between a diamagnetic band insulator and a paramagnetic Mott
insulator. Therefore, the ground state at the NIT may be metallic but
unconditionally
unstable to dimerization. Such a metallic behavior near the N-I borderline
has been indeed reported.\cite{saito} Depending on the strength of other 
interaction such as long range Coulomb interaction, electron-phonon (Peierls)
or electron-molecular-vibration (Holstein) couplings, continuous or 
discontinuous ionicity changes can be generated and the dimerization 
instability is more or less affected.\cite{anusooya}

Despite increasing computer power, first-principles calculations are still 
very rare in the field of molecular crystals. These calculations offer a 
unique tool for analyzing, at a microscopic level, CT salts without any 
beforehand assumption concerning the relative
strengths of electronic interactions.
They have already contributed to evidence the coupling between CT variation 
and anisotropic three-dimensional (3D) lattice contractions in the presence
of hydrogen bonds~\cite{oison} and to analyze the quantum intra and 
inter-chain interactions in the mixed stack CT crystal of 
TTF-2,5Cl$_2$BQ.\cite{katanjpcm} 
This compound has twice as less atoms per unit cell
as TTF-CA which makes the computational effort considerably lower, but only
few experimental work is available and structural data are limited to the 
ambient condition phase. From an artificial structure obtained by relaxing 
all atomic positions within the experimental unit cell, it has been shown 
that the weak molecular distortion due to symmetry breaking induces a 
non-negligible contribution to the total CT variation at the 
NIT.\cite{katanjpcm}

Starting from first-principles Density Functional Theory (DFT) 
calculations, we present here a thorough 
study of TTF-CA. Our main purpose is not to describe the phase 
transition itself but to reach a precise understanding of the 
electronic ground states on both sides and far from the transition point. 
This work gives a new and complementary insight for this family of
CT complexes. After a brief section concerning computational details, we
present, in Section~\ref{sec:tb}, 
a complete analysis of the valence and conduction bands with
the help of a tight-binding model fitted to the ab-initio results. 
This will lead us to discuss a possible mechanism of the phase transition
itself. In Section~\ref{sec:topo}, 
the 3D total electron density at 300 and 40 K will
be analyzed in details using Bader's topological approach.\cite{bader}
Section~\ref{sec:CT} will be devoted to the determination of 
the CT in the high and low temperature
phases from our tigh-binding model, from Bader's approach and from a
simple model based on isolated molecules calculations. 
The last method will only be briefly mentionned 
as we defer detailed description to a forthcoming publication.

\section{\label{sec:co}Computational details}

Our ab-initio ground state electronic structure calculations 
(frozen lattice)
are based on the experimental structures obtained
by neutron-scattering experiments at 300, 90 and 40~K.\cite{lecointe}
The symmetry is monoclinic with two equivalent chains with alternation of 
TTF and CA molecules along the stacking axis $\mathbf{a}$. 
This leads to 52 atoms per unit cell.
Above 81~K, the space group is $P12_1/n1$ and TTF and CA are located on
inversion centers. At 300~K : ${a}= 7.40$, ${b}= 7.62$, ${c}= 
14.59$~\AA\
and $\beta=99,1^\circ$ and at 90~K : ${a}= 7.22$, ${b}= 7.59$, 
${c}= 14.49$ \AA\ and $\beta=99,1^\circ$. At 40~K, the space group is $P1n1$
with ${a}= 7.19$, ${b}= 7.54$, ${c}= 14.44$ \AA\ and $\beta=98,6^\circ$.
TTF and CA are slightly distorded and displaced and form DA pairs in a 
ferroelectric arrangement in the \textbf{a} direction.

We performed electronic structure calculations within the framework of DFT 
using
the local density approximation (LDA) parametrization by Perdew 
and Zunger~\cite{pz}, 
Becke's gradient correction to the exchange energy~\cite{becke}
and Perdew's gradient correction to the correlation energy.\cite{perdew} 
The main difficulty in treating the molecular compounds of this family
is the presence of non negligible contributions of dynamical
electronic interactions of van der Waals type. They are known to
be poorly described within traditional DFT approximations, more
advanced treatments being far too time consuming for such complex
systems. Nevertheless, we believe that the presence of a significant
CT between D and A makes our ab-initio results reliable 
for the determination of the occupied electronic states,
especially
in the I phase where electrostatic interactions largely dominate.

We used the projector augmented wave (PAW) method~\cite{blo} 
which uses augmented 
plane waves to describe the full wave functions and densities without 
shape approximation. The core electrons are described within 
the frozen core approximation. 
The version of the CP-PAW code used for all our calculations considers
only $\mathbf{k}$ points where the wave functions are real.
In order to increase the number of $\mathbf{k}$ points
along $\mathbf{a}^\ast$, we had to double and treble the unit cell
along the $\mathbf{a}$ direction and our calculations showed that with 
3~$\mathbf{k}$ points along $\Gamma (0,0,0)\rightarrow 
\mathrm{X} (1,0,0)$ results are well
converged as for TTF-2,5Cl$_2$BQ.\cite{katanjpcm}
The band structures calculations have been
carried out with a plane wave cutoff of 30~Ry for the wave 
functions and 120~Ry for the densities, 
using 8~$\mathbf{k}$ points: $\Gamma (0,0,0)$,
$\mathrm{X}/2 (1/2,0,0)$, $\mathrm{X} (1,0,0)$, $\mathrm{Y} (0,1,0)$, 
$\mathrm{W}/2 (1,1/2,0)$, $\mathrm{W} (1,1,0)$, $\mathrm{Z} (0,0,1)$, and 
$\mathrm{S} (1,0,1)$ in units of $(\mathbf{a}^\ast/2,
\mathbf{b}^\ast/2,\mathbf{c}^\ast/2)$.

As the topological
analysis requires very accurate electron densities, we have first verified 
that results do not deteriorate when using only 
3~$\mathbf{k}$ points along $\Gamma \rightarrow \mathrm{X}$ instead of 8
and then increased the plane wave cutoff to 50~Ry for the wave functions.
The subsequent topological analysis was performed using the new
InteGriTy software package~\cite{integrity} which achieves topological
analysis following Bader's approach~\cite{bader} on electron
densities given on 3D grids. In order to obtain very accurate results,
a grid spacing close to 0.10~a.u has been choosen.

\section{\label{sec:tb}Ab-initio band structure}

Our ab-initio DFT calculations in the crystal provide  not only the
total electron density $\mathrm{n(\mathbf{r})}$ but also for each
band the energies and wave functions at all $\mathbf{k}$ points given 
in section~\ref{sec:co}. The present section is devoted to the sole
valence bands (VB) and conduction bands (CB) which are
directly related to the strong anisotropy observed experimentally in 
transport and excitation properties.\cite{tokura,mitani,okamoto}
The detailed analysis of $\mathrm{n(\mathbf{r})}$ will be given in
section~\ref{sec:topo}.

\subsection{Quasi 1D shape of valence and conduction
bands}

\begin{figure}
\centerline{
\resizebox{8.6cm}{!}{
\rotatebox{-90}{\includegraphics*[1.0cm,1.7cm][20cm,21.6cm]{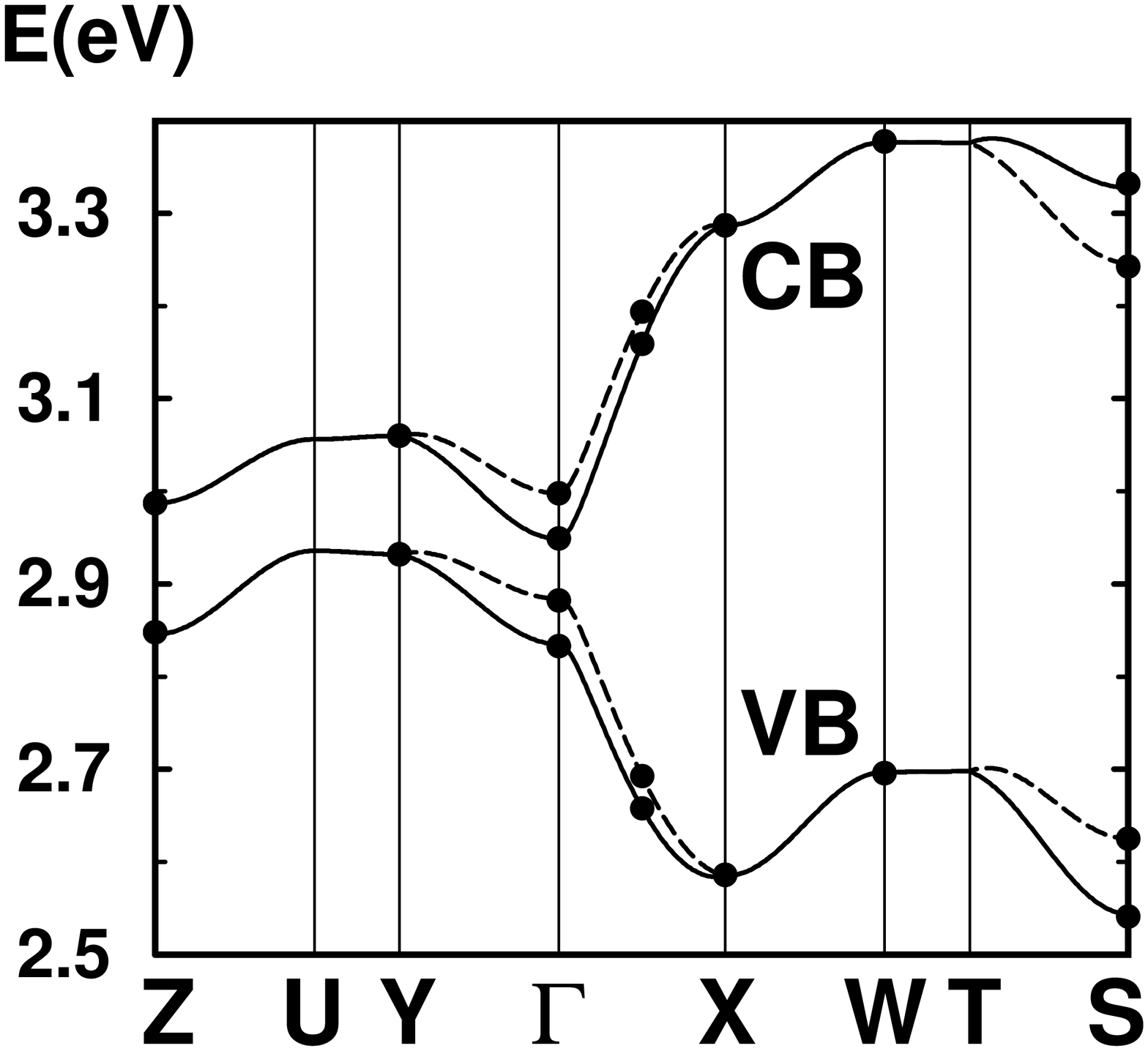}}
\hspace{0.2cm}
\rotatebox{-90}{\includegraphics*[1.0cm,3.5cm][20cm,23.7cm]{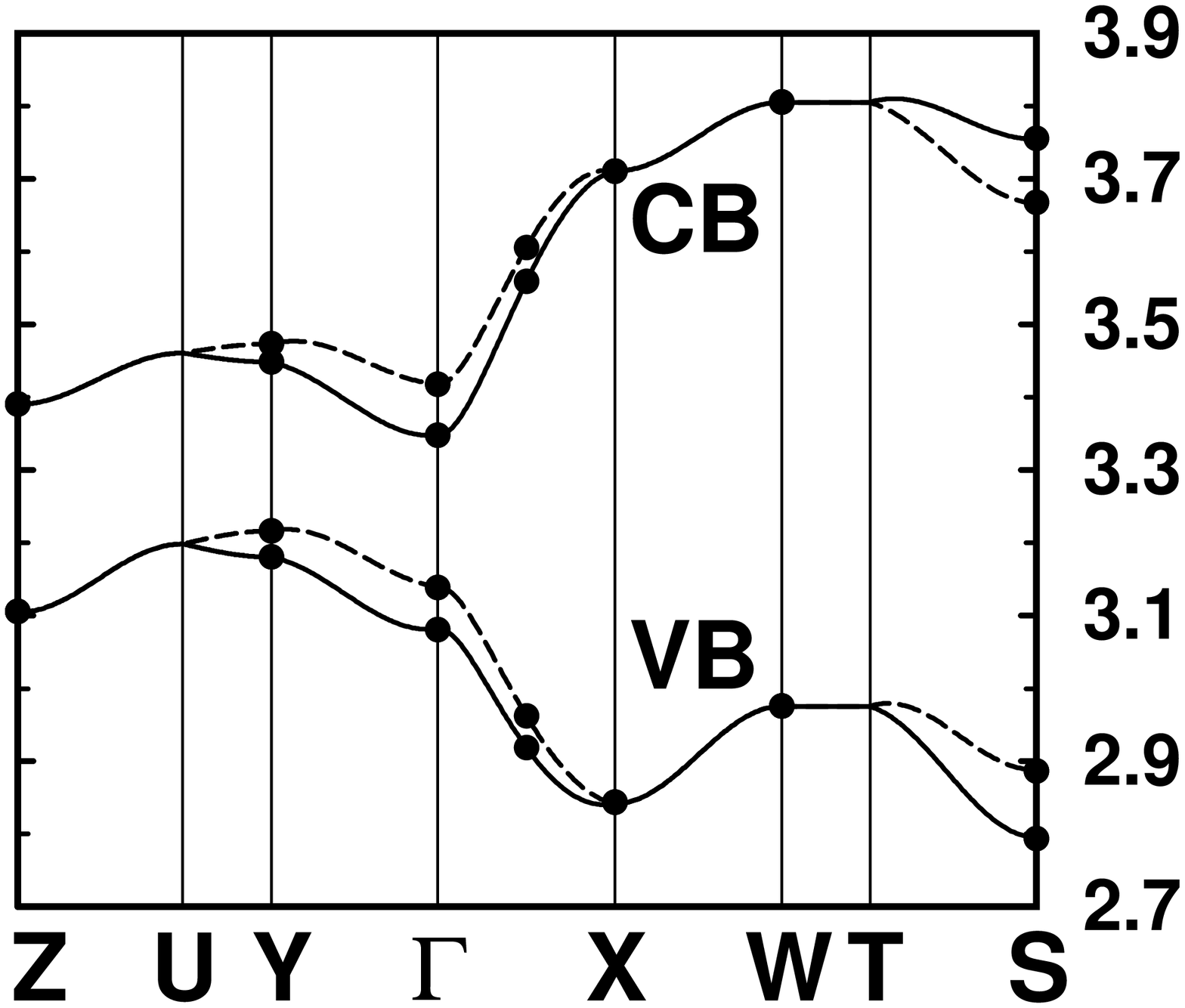}}
}}
\centerline{\hfill 300~K \hfill 40~K\hfill}
\caption{\label{disp} Valence and conduction bands of TTF-CA in the
high temperature phase (300~K, on the left) and low temperature phase
(40~K, on the right). Dots correspond to the ab-initio values and 
curves to the tight-binding model presented in Section~\ref{sec:tb}.
}
\end{figure}

The dispersion curves for the VB and CB
are given in Fig.~\ref{disp} for the experimental
structures at 300~K (high temperature phase) and 40~K 
(low temperature phase). In both cases, they are separated 
by more than 1~eV from the other occupied and unoccupied bands forming 
thus a nearly isolated four-band system. The presence of two equivalent
chains in the unit cell related by a gliding plane makes the VB and CB 
twofold, splittings in some particular directions
being due to small interactions between the symmetry related
chains. The dispersion is maximum along $\Gamma \rightarrow \mathrm{X}$ 
which is the direction of reciprocal space corresponding to the chain axis 
$\mathbf{a}$.\cite{katanssc}

\begin{figure*}
\centerline{
\resizebox{16cm}{!}{
{\includegraphics{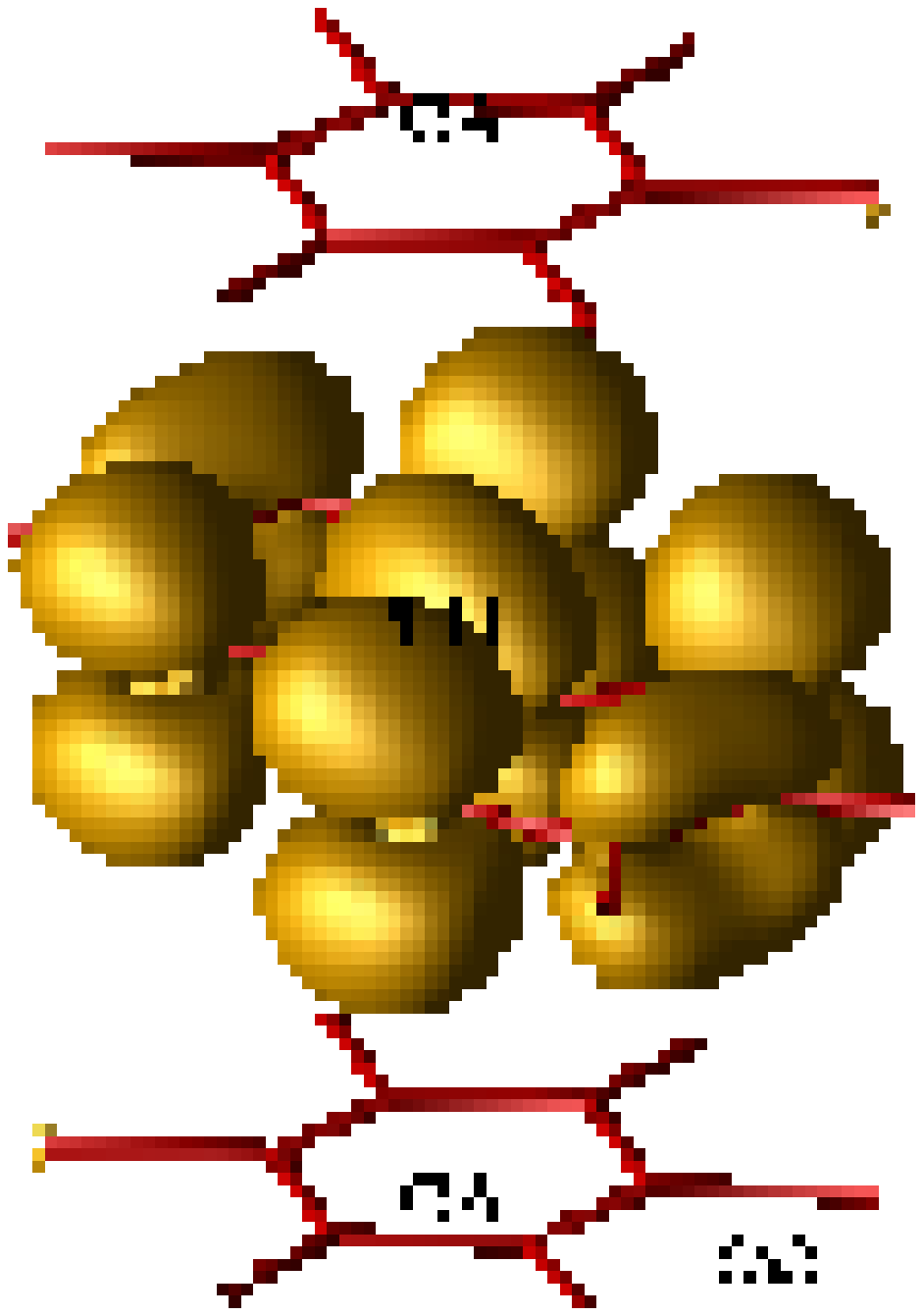}}
\hspace{2.cm}
{\includegraphics{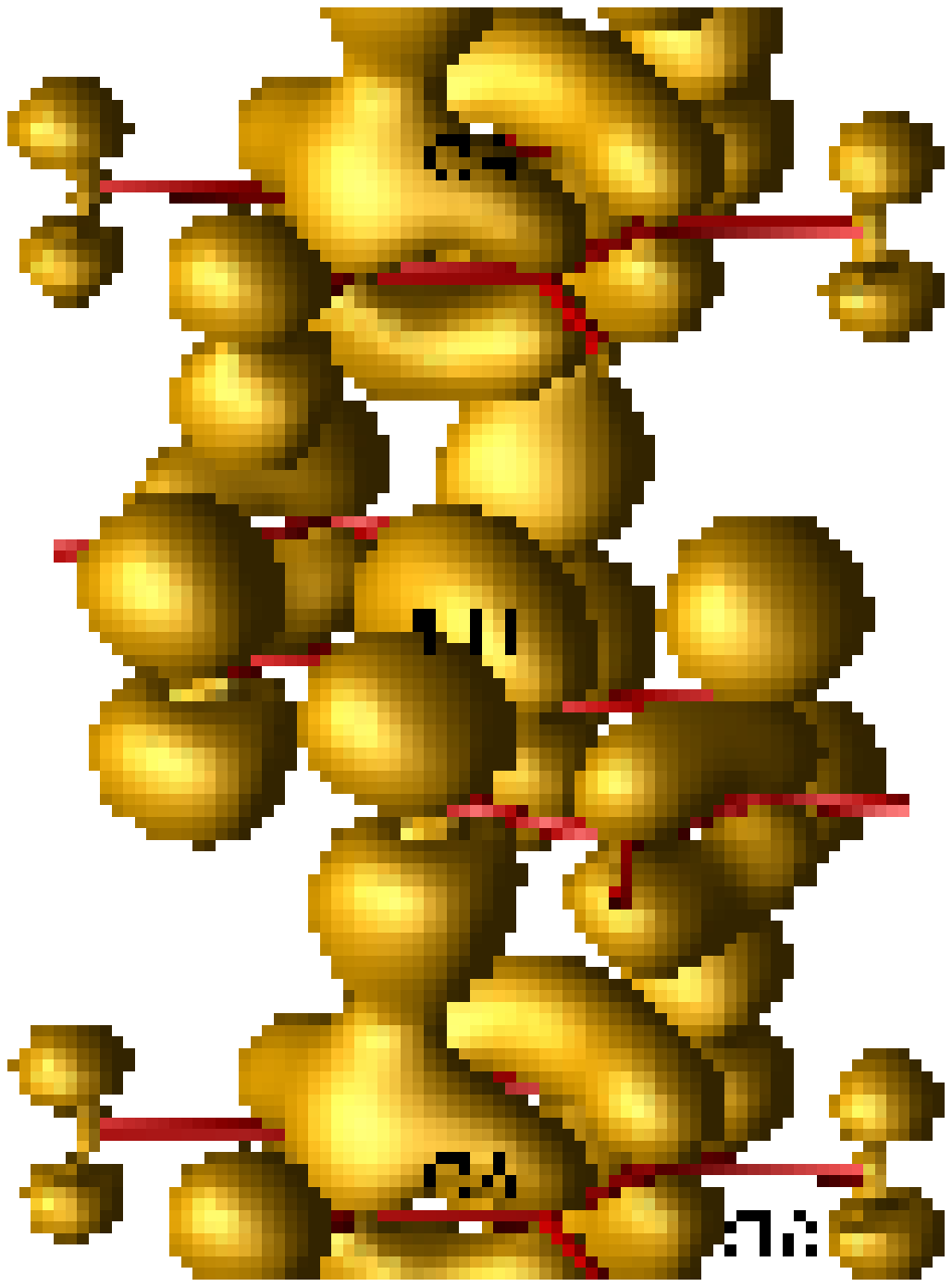}}
\hspace{2.2cm}
{\includegraphics{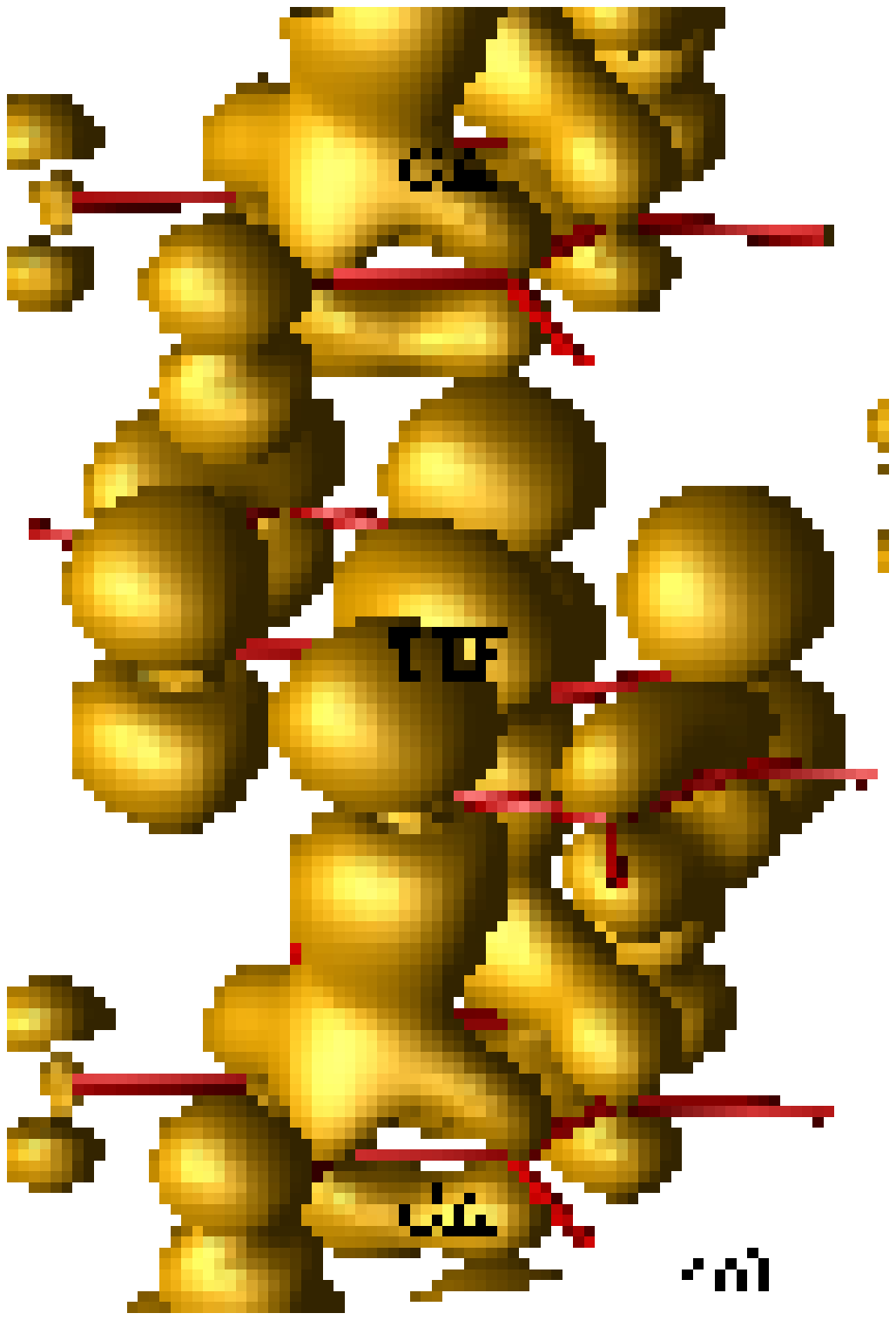}}
}}
\vspace{0.2cm}
\centerline{\hfill $\Gamma$ at 300~K \hfill $\mathrm{X}$ 
at 300~K \hfill $\Gamma$ at 40~K \hfill}
\caption{\label{isoVB} Isodensity representation of the valence band 
states for one chain (contours at 0.0004~e$^-/$a.u.$^{3}$) of TTF-CA:  
(a) for $\mathbf{k}=\Gamma$ (left) and (b) for $\mathbf{k}=\mathrm{X}$
(middle) in the high temperature phase (300~K)
and (c) for $\mathbf{k}=\Gamma$ (right)
in the low temperature phase (40~K).
}
\end{figure*}

The isodensity representation of the VB states are given in Fig.~\ref{isoVB}.
At 300~K for $\mathbf{k}=\Gamma$ (Fig.~\ref{isoVB}a), the density is located 
on TTF with a shape very close to that of the highest occupied molecular 
orbital (HOMO) of TTF.\cite{katanTTF} At $\mathbf{k}=\mathrm{X}$, 
(Fig.~\ref{isoVB}b), hybridization occurs and the density centered 
on the acceptor molecule is similar 
to the one of the lowest unoccupied molecular orbital (LUMO)
of CA.\cite{katanCA} 
The same happens in TTF-2,5Cl$_2$BQ~\cite{katanssc}
and is a direct consequence of the presence of an inversion center
in the crystal combined to different symmetries of the molecular orbitals
which are involved. 
As the inversion center is lost at 
40~K, hybridization is no more symmetry forbidden at 
$\mathbf{k}=\Gamma$ and
the pairing of TTF and CA is clearly evidenced in Fig.~\ref{isoVB}c
(DA pair on the bottom). On
these figures one can identify the regions of wave function overlap
responsible for the intra-chain hopping integrals.

The dispersion is much less important along 
$\Gamma \rightarrow \mathrm{Y}$, 
nearly vanishing in the third direction and very similar for both 
structures. As for TTF-2,5Cl$_2$BQ,\cite{katanjpcm} the band gap is
most probably indirect: $E_g = 0.05~\mbox{eV}$ at 300~K and $0.15~\mbox{eV}$ 
at 40~K.  It is well known that the LDA used in our calculations
underestimate the gap of semiconductors; these values are thus
lower estimates of the true band gap. However, because of missing data a
direct comparison to experiments is impossible. Optical absorption
measurements~\cite{torrance} showed a first broad peak around 0.7~eV for the
N phase which is attributed to the ionization of one DA pair. It may 
correspond either to band to band transitions 
or to an exciton peak. The optical 
spectra of TTF-2,5Cl$_2$BQ~\cite{torrance} shows a shift to higher energies
which is coherent with a larger
calculated band gap for TTF-2,5Cl$_2$BQ: 0.14~eV in the high 
temperature structure.\cite{katanjpcm} 

\subsection{\label{subsec:tb}Tight-binding model} 

\begin{figure}
\centerline{
\resizebox{8.6cm}{!}{
\rotatebox{-90}{\includegraphics*[1.0cm,1.5cm][20.5cm,21.6cm]{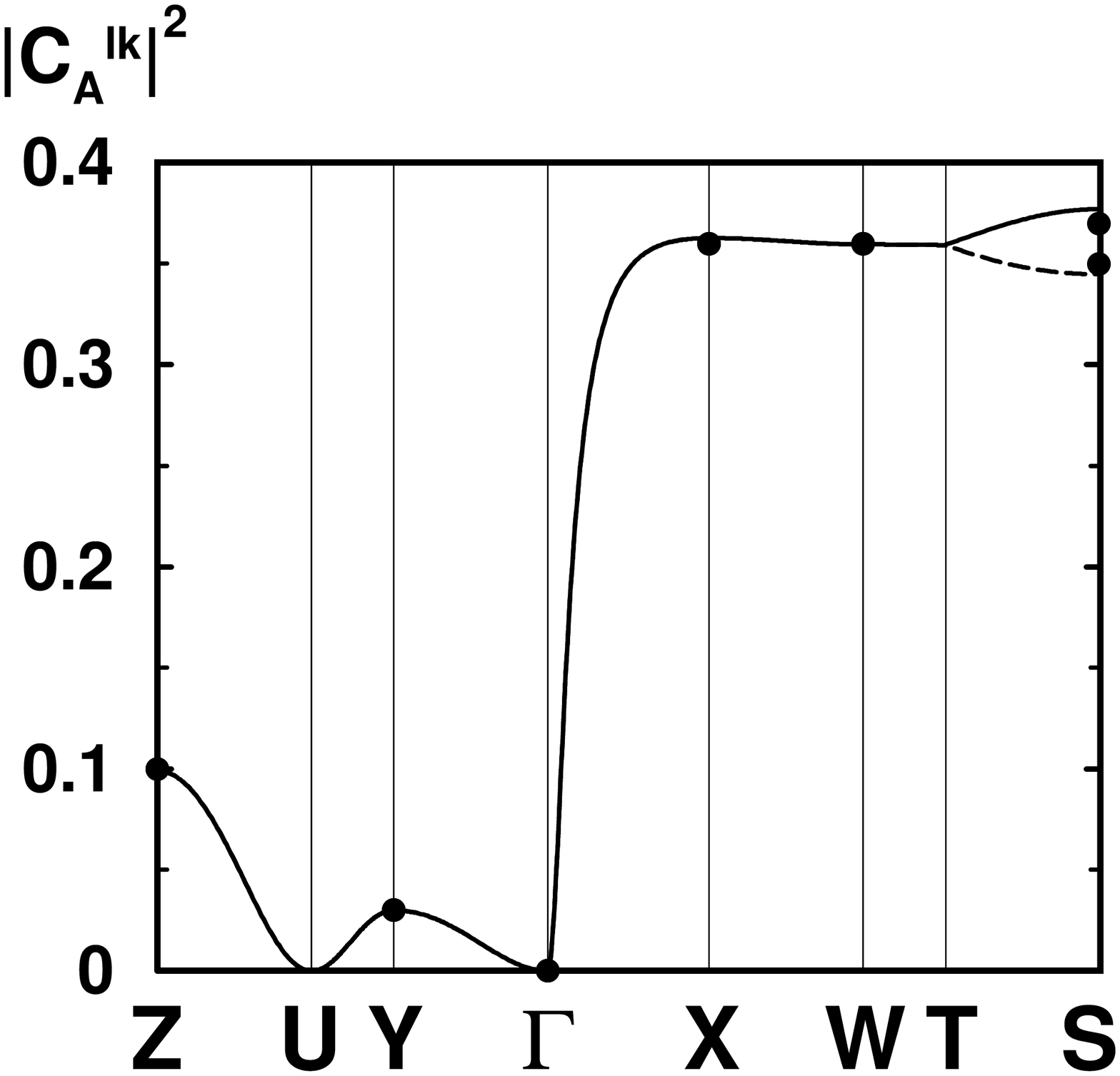}}
\hspace{0.2cm}
\rotatebox{-90}{\includegraphics*[1.0cm,3.5cm][20.5cm,23.7cm]{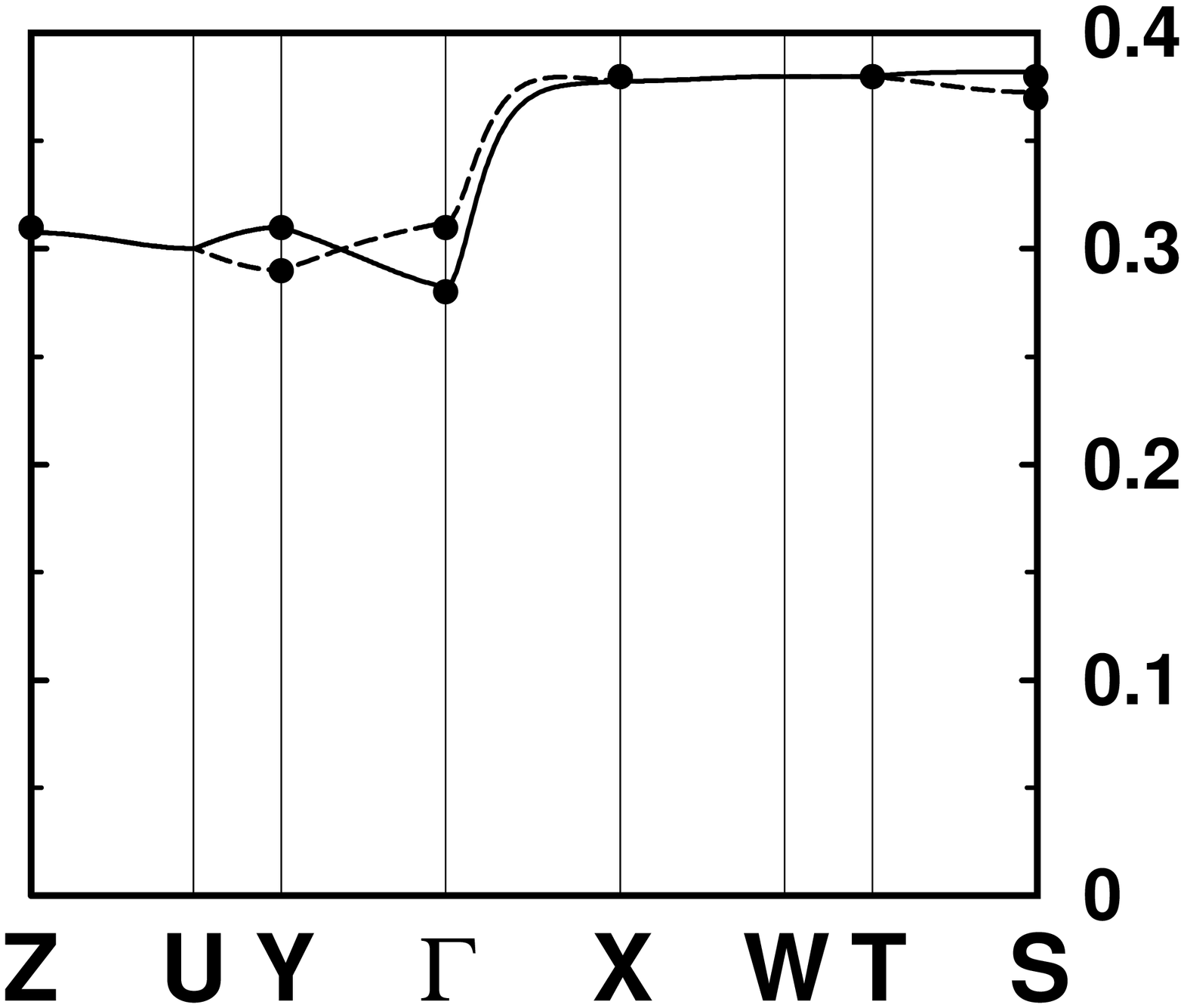}}
}}
\centerline{\hfill 300~K \hfill 40~K\hfill}
\caption{\label{Cak2} Weights on the acceptor $\vert\mathrm{C^{lk}_A}\vert^2$
of the occupied valence states in the
high temperature phase (300~K, on the left) and low temperature phase
(40~K, on the right). Dots correspond to the ab-initio values and 
curves to the tight-binding model.
}
\end{figure}

As shown in Fig.~\ref{isoVB}, the ab-initio VB and CB can be interpreted in
the frame of a tight-binding model as linear combinations
of the HOMO of TTF and the LUMO of CA. 
One should emphasize that in such a model, hybridizations with lower
bands are completely neglected. 
The interaction with deeper occupied states is nevertheless
indirectly taken into account via the values of the parameters,
which are fitted to the VB and CB given by an ab-initio calculation
of all states.
We will follow a similar procedure to the
one used for TTF-2,5Cl$_2$BQ,\cite{katanjpcm} but here the task is more
involved as we have to handle four bands instead of two. Let us write
$\mathrm{\vert{\mathbf{r},D}\rangle}$ and 
$\mathrm{\vert{\mathbf{r'},A}\rangle}$ respectively the HOMO of TTF 
located at position $\mathbf{r}$ and the LUMO of CA at
$\mathbf{r'}$. The VB Bloch functions are then defined by:
\begin{eqnarray}
\mathrm{\vert\Psi^{lk}\rangle\!\!\!\!}&& 
\mathrm{= \sum_{n}e^{i\mathbf{k}\mathbf{R_n}}\biggl(
C^{lk}_D\vert{\mathbf{R_n},D}\rangle + C^{lk}_{D'}
\vert{\mathbf{R_n}+\mathbf{\nu},D'}\rangle}  
\nonumber\\
&&
\mathrm{+ \; C^{lk}_A \vert{\mathbf{R_n}+\mathbf{\tau},A}\rangle 
+ C^{lk}_{A'} \vert{\mathbf{R_n}+\mathbf{\tau}+\mathbf{\nu},A'}
\rangle\biggr)}.
\label{eqpsi}
\end{eqnarray}
Here $\mathrm{l}=1,2$ is the band index; $\mathbf{R_n}$ corresponds to a
primitive translation; $\mathbf{\tau} = \mathrm{\mathbf{a}/2}$ 
represents a translation between A and D along the chain axis and 
$\mathbf{\nu} = \mathrm{\mathbf{a}/2+\mathbf{b}/2+\mathbf{c}/2}$ 
a translation between the two symmetry equivalent chains, the
molecules from the second chain being denoted by $\mathrm{D'}$
and $\mathrm{A'}$.
The molecular orbitals $\mathrm{\vert{D}\rangle}$, 
$\mathrm{\vert{A}\rangle}$,
$\mathrm{\vert{D'}\rangle}$ and $\mathrm{\vert{A'}\rangle}$
are calculated using ab-initio simulations for isolated
molecules in their experimental crystalline conformations. 
For each $\mathbf{k}$ point, the weight 
$\vert\mathrm{C^{lk}_j}\vert^2$ ($\mathrm{j=}$ A or D) 
is obtained by mean square minimization in 
the unit cell of the difference between the ab-initio value of
$\mathrm{\vert\Psi^{lk}(\mathbf{r})\vert^2}$ 
calculated in the crystal and the one deduced
from expression~(\ref{eqpsi}). 
Fig.~\ref{Cak2} displays these weights at the 6 corners of the reduced 
Brillouin zone used in these calculations for both the high (left) 
and low (right) temperature 
phases. In the former, the valence band at the $\Gamma$ is 
a pure TTF state for symmetry reasons: $\mathrm{C_A^{l\Gamma}}=0$. 
The band gap separates nearly pure D occupied states 
from nearly pure A empty states in the whole 
$\mathrm{Z\!-\!U\!-\!Y\!-}\Gamma$
plane, whereas electronic states are of mixed D-A nature in the
$\mathrm{X\!-\!W\!-\!T\!-\!S}$ plane. In the latter, a significant 
mixing of D-A states takes place in the whole Brillouin zone, for
the VB as well as for the CB.

The charge transfer $\mathrm{\rho_{VB}}$
from TTF to CA is given by the mean value of the weights of the
acceptors in the VB:
\begin{eqnarray}
\mathrm{\rho_{VB}}
&&
\mathrm{= \frac{1}{2}\sum_{l}\biggl(
\langle\vert C^{lk}_A    \vert^2 \rangle +
\langle\vert C^{lk}_{A'} \vert^2 \rangle \biggl)}
\nonumber\\
&&
\mathrm{= \frac{V}{8\pi^3} \int_{ZB}\sum_l |C_A^{lk}|^2d^3\mathbf{k}}.
\label{eqrhovb}
\end{eqnarray}
Fig.~\ref{Cak2} shows clearly that the increase of $\mathrm{\rho_{VB}}$
in the low temperature phase results mainly from additionnal hybridization,
which is no longer symmetry forbidden, in the whole 
$\mathrm{\Gamma\!-\!Y\!-\!U\!-\!Z}$ plane. 

\begin{figure}
\centerline{
\resizebox{7.0cm}{!}{
{\includegraphics{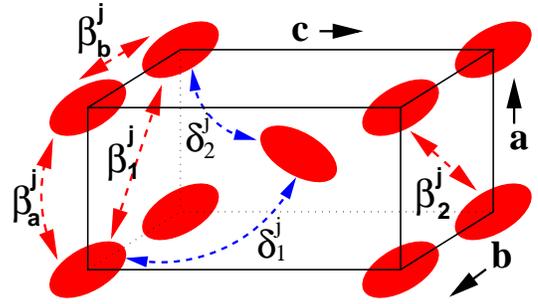}}
}}
\caption{\label{parajj} 
Definition of the interaction parameters between two molecules
(denoted by j, j$=$ D or A) of the same species.
}
\end{figure}

\begin{figure}
\centerline{
\resizebox{7.0cm}{!}{
{\includegraphics{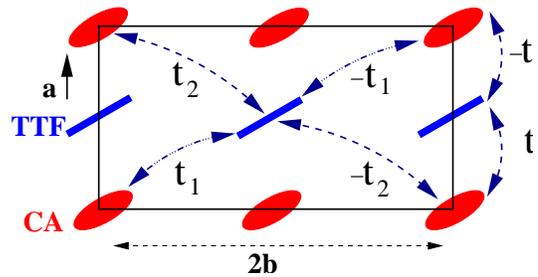}}
}}
\caption{\label{paraDA-ab} 
Definition of the interaction parameters between TTF and
CA in the $\mathrm{(a,b)}$ plane. $\mathrm{\pm t}$ is the hopping 
integral along the $\pi$ chains. The orientation of the molecules
in the chains parallel to the $\mathrm{\mathbf{a}/2+\mathbf{b}}$
and $\mathrm{-\mathbf{a}/2+\mathbf{b}}$ being different, 
the hopping integrals $\mathrm{t}_1$ and $\mathrm{t}_2$ are also
different.
}
\end{figure}

\begin{figure}
\centerline{
\resizebox{6.0cm}{!}{
{\includegraphics{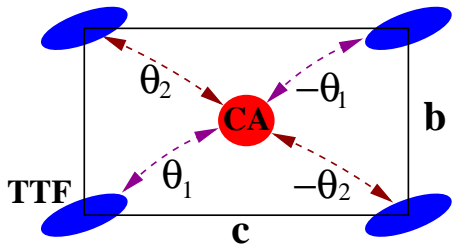}}
}}
\caption{\label{paraDA-pipi} 
Definition of the interaction parameters between between TTF and
CA in the $\mathrm{(b,c)}$ plane.
}
\end{figure}

In order to model VB and CB in terms of a tight-binding scheme, we
define the on-site energies by 
$\mathrm{\langle D \vert H \vert D \rangle}
\mathrm{= \langle D' \vert H \vert D' \rangle}
\mathrm{= E_0^D}$ and
$\mathrm{\langle A \vert H \vert A \rangle}
\mathrm{= \langle A' \vert H \vert A' \rangle}
\mathrm{= E_0^A.}$
Two families of intermolecular interactions have been defined,
the first corresponding to interactions between molecules of
the same species shown on Fig.~\ref{parajj}
and the second to interaction between TTF and CA
shown on Fig.~\ref{paraDA-ab} and  Fig.~\ref{paraDA-pipi}. 
Due to the different symmetries of the HOMO of TTF and LUMO
of CA, the sign of the hopping integrals between TTF and CA
is alternately plus and minus. For example, along the stacking chains
one has $\langle \mathbf{0},\mathrm{D \vert H} \vert \mathbf{a}/2,\mathrm{A} 
\rangle = \mathrm{t}$ and $\langle \mathbf{a}/2,\mathrm{A \vert H} \vert
\mathbf{a}, \mathrm{D} \rangle = - \mathrm{t}$.
The symmetry breaking in the low temperature phase leads to additionnal
terms so that the latter hopping integrals become respectively
$\mathrm{t} + \varepsilon$ and $-\mathrm{t} + \varepsilon$ 
whereas $\mathrm{\pm t_{i}}$, $\mathrm{\pm \theta_{i}}$ and
$\mathrm{\delta_i^j}$ ($\mathrm{i = 1}$ or $2$ and $\mathrm{j= D}$ 
or $\mathrm{A}$) become $\mathrm{\pm t_{i} + \varepsilon_i}$,
$\mathrm{\pm \theta_{i} + \lambda_i}$ and $\mathrm{\delta_i^j \pm 
\upsilon_i^j}$ respectively. 

Using the ab-initio energies and the weights 
$\mathrm{\vert C^{lk}_j \vert^2}$  at the 6 corners
of the reduced Brillouin zone used in these calculations, 
we determine the parameters given on
Table~\ref{tab:jj} to \ref{tab:def} and the curves plotted in 
Fig.~\ref{disp} and Fig.~\ref{Cak2}. 
Some parameters like $\mathrm{\beta_1^j}$
 and $\mathrm{\beta_2^j}$ (Fig.~\ref{parajj})
cannot be determined separately because 
of the small number of $\mathbf{k}$ points available, but all 
these contributions are found to be negligible. In Fig.~\ref{disp} 
the degeneracy is lifted at $\Gamma$ due to the small inter-chain 
interactions $\mathrm{\delta_i^j}$ (Fig.~\ref{parajj}), 
at $\mathrm{S}$ due to 
$\mathrm{\theta_i}$ (Fig.~\ref{paraDA-pipi}) and, 
in the low temperature phase at
$\mathrm{Y}$ due to $\mathrm{\upsilon_i^j}$.
The $\mathrm{\theta_i}$ are also responsible for the small HOMO-LUMO
hybridization at $\mathrm{Y}$ and $\mathrm{Z}$ (Fig.~\ref{Cak2}).

The transfer integral $\pm\mathrm{t}$ along the chain axis 
(Fig.~\ref{paraDA-ab})
dominates and its estimate is in agreement with previous 
ones.\cite{jacobsen,girlando1,tokura, katanssc} 
In the low temperature phase,  $\mathrm{t}$ 
increases significantly (Table~\ref{tab:DA}) 
partly due to the lattice contraction and partly to the symmetry 
breaking.\cite{katanjpcm} 
As for TTF-2,5Cl$_2$BQ, the main effect of the latter is expressed by
the additional parameter $\varepsilon$ which is as large as a
third of $\mathrm{t}$ (Table~\ref{tab:def}) and which leads to a
large HOMO-LUMO hybridization in the
$\mathrm{Z\!-\!U\!-\!Y\!-}\Gamma$ plane. The other additional
parameters $\mathrm{\varepsilon_i}$, $\mathrm{\lambda_i}$ 
and $\mathrm{\upsilon_i^j}$ have negligible values.

\begin{table}
\caption{\label{tab:jj} 
On site energies and interaction parameters (eV) between two molecules
of the same species as defined in Fig.~\ref{parajj}.}
\begin{ruledtabular}
\begin{tabular}{lrrlrr}
&300~K&40~K&&300~K&40~K\\
\hline
$\mathrm{E_0^D}$ & 2.894 & 3.191 & $\mathrm{E_0^A}$ & 3.048 & 3.348 \\
$\mathrm{\beta^D_a}$ & 0.002 & 0.001 & $\mathrm{\beta^A_a}$ & $-$0.017
 & $-$0.024 \\
$\mathrm{\beta^D_b}$ & $-$0.022 & $-$0.005 & $\mathrm{\beta^A_b}$ & $-$0.023
&  $-$0.003 \\
$\mathrm{\beta^D_1 + \beta^D_2}$ & 0.003 & 0.004 & $\mathrm{\beta^A_1
  + \beta^A_2}$ & 0.002 & 0.003 \\
$\mathrm{\delta^D_1}$ & $-$0.002 & $-$0.003 & $\mathrm{\delta^A_1}$ &
$-$0.004 & $-$0.004 \\ 
$\mathrm{\delta^D_2}$ & $-$0.004 & $-$0.005 & $\mathrm{\delta^A_2}$ &
$-$0.002 &  $-$0.003 \\
\end{tabular}
\end{ruledtabular}
\end{table}

\begin{table}
\caption{\label{tab:DA} 
Interaction parameters (eV) between TTF and CA as defined in 
Fig.~\ref{paraDA-ab} and Fig.~\ref{paraDA-pipi}.
}
\begin{ruledtabular}
\begin{tabular}{lcc}
& 300~K&40~K\\
\hline
$\mathrm{t}$ & 0.167 & 0.206 \\
$\mathrm{t_1 -t_2}$ & 0.003 & 0.005 \\
$\mathrm{\theta_1}$ & 0.016 & 0.012 \\
$\mathrm{\theta_2}$ & 0.005 & 0.010 \\
\end{tabular}
\end{ruledtabular}
\end{table}

\begin{table}
\caption{\label{tab:def} 
Additionnal deformation parameters (eV) due to the symmetry
breaking in the low temperature phase (40~K) as defined in 
Section~\ref{subsec:tb}.
}
\begin{ruledtabular}
\begin{tabular}{ccccc}
$\mathrm{\varepsilon}$ & $\mathrm{\varepsilon_1 +
  \varepsilon_2}$ & $\mathrm{\lambda_1 + \lambda_2}$ &
$\mathrm{\upsilon^D_1 + \upsilon^D_2}$ & $\mathrm{\upsilon^A_1 +
  \upsilon^A_2}$ \\
\hline
0.061 & 0.001 & -0.002 & -0.003 & -0.004 \\
\end{tabular}
\end{ruledtabular}
\end{table}

\subsection{Towards a metallic state?} 

\begin{figure}
\centerline{
\resizebox{5.0cm}{!}{
\rotatebox{-90}{\includegraphics*[3.0cm,1.7cm][20cm,24.cm]{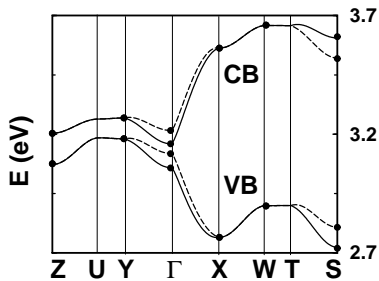}}
}}
\caption{\label{disp90} Valence and Conduction bands of TTF-CA for the
experimental structure at 90~K. Dots correspond to the ab-initio values 
and curves to the tight-binding model presented in Section~\ref{sec:tb}.
}
\end{figure}

Our calculations at 300 and 40~K
present the electronic structure far from
the transition. In order to understand how the band structure is
changed by thermal contraction, we performed also a calculation
just above the NIT for the crystal structure known at 90~K.
The corresponding dispersion curves
are given on Fig.~\ref{disp90}. They display a closed indirect band 
gap between $\Gamma$ and $\mathrm{Y}$, with nearly pure D occupied 
states at $\mathrm{Y}$ at a higher energy than nearly pure A empty 
states at $\Gamma$, as our method 
imposes a fully occupied VB. This is of course unphysical and is due 
to the DFT-LDA error on the band gap value. It leads however to 
interesting conclusions: (i) the thermal lattice contraction 
reduces the indirect band gap in the neutral phase, going toward a 
metallic state and (ii) as this band gap closes, an electron transfer
occurs necessarily from D to A in the vicinity of the transition
temperature, leading to an increase of the ionicity of the 
molecules, even in a frozen lattice.

Within our tight-binding model, the band gap 
decrease results from (i) the decrease of the difference 
$\mathrm{E_0^A-E_0^D}$ between
the on site energies,
(ii) the increase of the interaction between two TTF molecules
along $\mathbf{b}$ ($\mathrm{\vert\beta_b^D\vert}$) which increases
the energy of the valence band at $\mathrm{Y}$ and (iii) the 
increase of the interaction between the two CA molecules along
$\mathbf{a}$ ($\mathrm{\vert\beta_a^A\vert}$) which decreases
the energy of the CB at $\Gamma$ (Table~\ref{tab:90}).
The cell contraction increases also $\mathrm{t}$ resulting in a slight 
increase of the weights of the acceptors in the VB
$\mathrm{\vert C_A^{lk}\vert^2}$ in the $\mathrm{X\!-\!W\!-\!T\!-\!S}$
plane from 0.36 to 0.38 and a very small variation of $\mathrm{\rho_{VB}}$
as observed experimentally~\cite{jacobsen} before the transition. 
One can notice that 
 $\mathrm{\vert C_A^{lk}\vert^2}$ in the $\mathrm{X\!-\!W\!-\!T\!-\!S}$
plane has already reached at 90~K the value that we obtained at 40~K.

Of course, in the actual crystal, the charge transfer is largely
facilitated by dynamical effects in the vicinity of the transition
temperature, when the gap becomes narrow. Thermal 1D excitations
from D to A have already been observed.\cite{mhl} They are 
precursors of the global changes in the crystal which certainly
involve intra and intermolecular vibrations.
Calculations on a A-D-A model complex~\cite{oison} have recently 
shown that the CT variation occuring in the VB (along the
mixed stack $\pi$-chains) induces an intra-molecular electronic
redistribution, affecting deep molecular states as far as
10~eV below the frontier orbitals. These deep states are also
concerned in the hydrogen bonds (along OH-chains resulting from 
coplanar alternating D and A molecules in the $\mathbf{b} \pm 
\mathbf{a}/2$ direction)
so that their modifications are directly related to the lateral 
lattice contractions.
In the low temperature phase, the gap opens again between mixed D-A
states both in the VB and CB, in the whole Brillouin zone.
This suggest an electronic mechanism for the NIT where the electronic 
redistribution associated to the structural phase transition allows the
crystal to avoid the metallic state.\cite{anusooya} 

\begin{table}
\caption{\label{tab:90} 
Parameters of the tight-binding model (eV) mostly affected by the
structural changes occuring between 300 and 90~K.
}
\begin{ruledtabular}
\begin{tabular}{lrr}
& 300~K & 90~K \\
\hline
$\mathrm{E_0^A - E_0^D}$ & 0.154 & 0.135 \\
$\mathrm{\beta^D_b}$ & $-$0.022 & $-$0.028  \\
 $\mathrm{\beta^A_a}$ & $-$0.017 & $-$0.022 \\
$\mathrm{t}$ & 0.167 & 0.189 \\
\end{tabular}
\end{ruledtabular}
\end{table}

\section{\label{sec:topo}Intermolecular interactions}

The previous section showed that VB
and CB are clearly 1D. The aim of the present section is to 
explore thoroughly the total electron density $\mathrm{n(\mathbf{r})}$ 
given by our first-principle calculations. 
This is essential to elucidate all features, not only the
1D aspects, presented by this NIT.

We could plot 
isodensity representations of $\mathrm{n(\mathbf{r})}$ in some selected
planes to evidence the most important intermolecular couplings. 
But, the presence of two chains per unit cell related by a
gliding plane makes it an endless task and Bader's
approach~\cite{bader} is a clever manner to circumvent this
difficulty.

\subsection{Bader's theory}

Within the Quantum Theory of Atoms in Molecules~\cite{bader}
$\mathrm{n(\mathbf{r})}$ can be analysed in details by means of its 
topological properties. The topological features of 
$\mathrm{n(\mathbf{r})}$
are characterized by analysing its gradient vector field
$\mathbf{\nabla}\mathrm{n}(\mathbf{r})$. Here we will focus on
bond critical points (CP) where $\mathbf{\nabla}\mathrm{n}(\mathbf{r})$
vanishes. They are characterized by the density at the CP 
$\mathrm{n(\mathbf{r}_{CP})}$, a positive curvature ($\lambda_3$)
parallel and two negative curvatures ($\lambda_1,\lambda_2$)
perpendicular to the bond path. The ellipticity of a bond, defined as 
$\epsilon = \lambda_1/\lambda_2 - 1$,
describes the deviation from cylindrically symmetric bonds.
For closed shell interactions, further information is 
obtained by using Abramov's~\cite{abramov} expression for the 
kinetic energy density $\mathrm{G(\mathbf{r}_{CP})= 
3/10 (3\pi^2)^{2/3} n(\mathbf{r}_{CP})^{5/3} 
+ 1/6 \nabla^2 n(\mathbf{r}_{CP})}$, and, combined 
with the local form of the Virial theorem~\cite{bader}, the 
local contribution to the potential energy density 
$\mathrm{V(\mathbf{r}_{CP})= 1/4 \times \nabla^2 n(\mathbf{r}_{CP})
- 2 G(\mathbf{r}_{CP})}$ which have already been used to study
hydrogen bond strengths.\cite{espinosa} In the present work, 
$\mathrm{V(\mathbf{r}_{CP})}$ will be used to quantify the strength
(intensity) of the intermolecular
interactions. It allows to quantitatively compare  
interactions between two or more atoms. This procedure is
less speculative than the comparison of interatomic distances to 
the sum of van der Waals radii~\cite{lecointe},
or related criteria~\cite{zefirov,collet}, as van der Waals radii
are not defined with sufficient accuracy~\cite{pauling,bondi,baur}
and fail to take into account different relative orientations.

\subsection{\label{sub:intra}Intra-chain interactions}

\begin{figure*}
\resizebox{14cm}{!}{
{\includegraphics{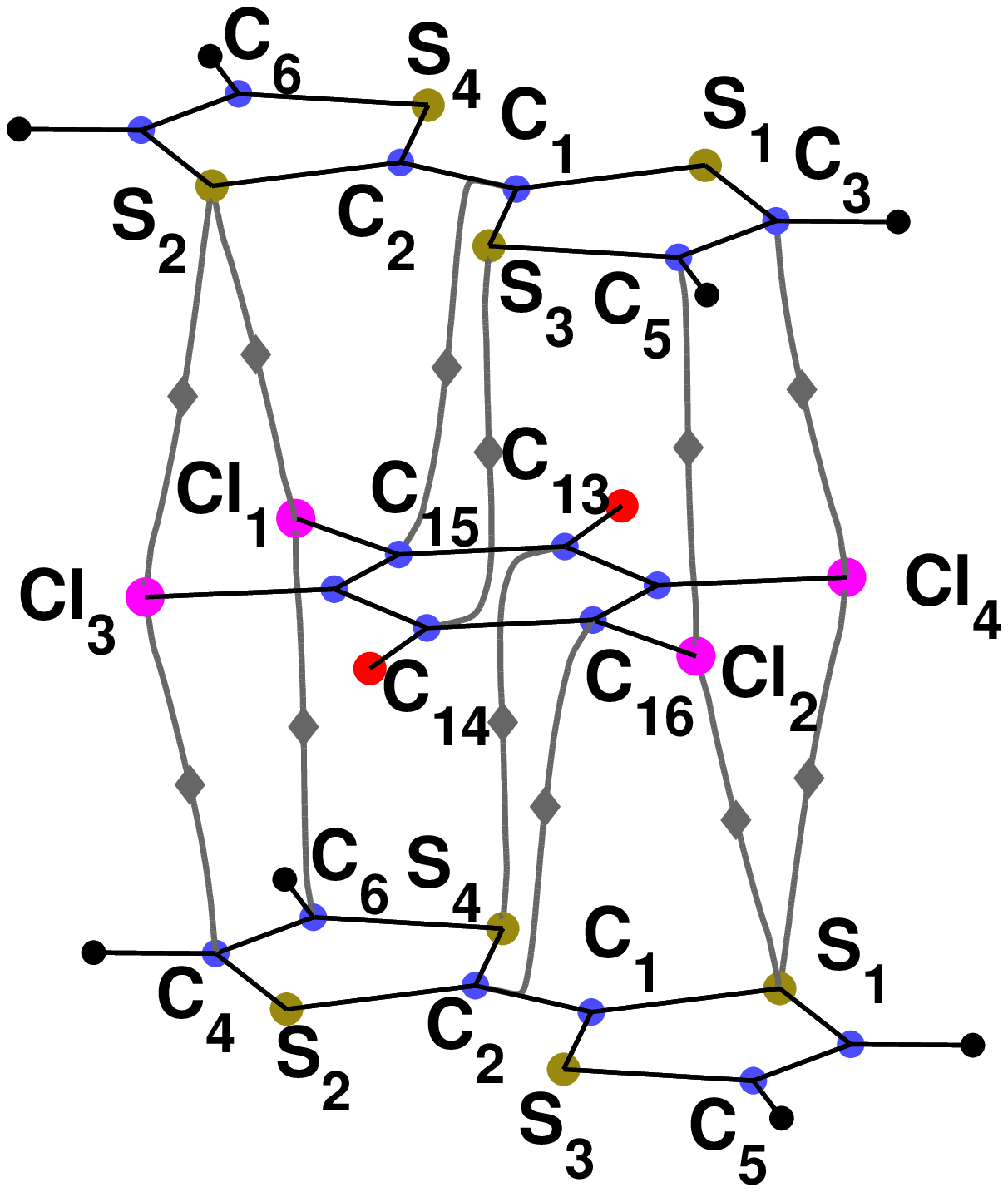}}
{\includegraphics{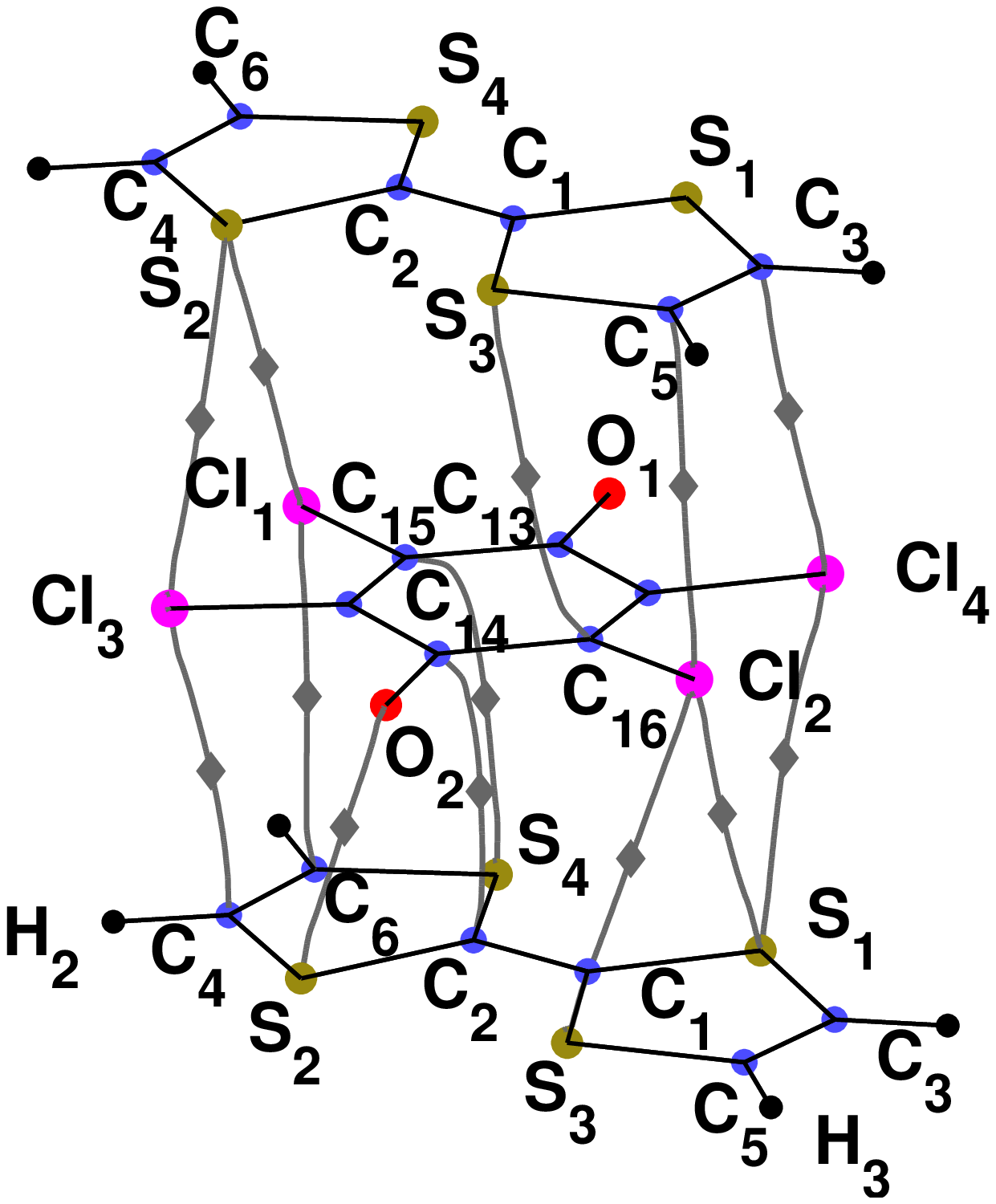}}
}
\vspace{0.2cm}
\centerline{\hfill 300~K \hfill 40~K\hfill}
\caption{\label{cpintra} Intra-chain interactions in the
high temperature phase (300~K, on the left) and low temperature phase
(40~K, on the right). Diamonds correspond to the bond critical points 
and lines indicate the bond path associated to each critical point.
}
\end{figure*}

In the high temperature phase (300~K), the strongest contact within
the chains is the one connecting $\mathrm{S_4}$ to 
$\mathrm{C_{13}}$-$\mathrm{C_{15}}$ (Fig.~\ref{cpintra} and 
Table~\ref{tab:cpintra}). It corresponds to the one already visible
in the VB (Fig.~\ref{isoVB}) and has thus a significant contribution 
resulting from HOMO-LUMO overlap. It has a quite large ellipticity
which is related to the fact that interaction do not occur between 
two atoms but rather between $\mathrm{S_4}$ and the bridge of
$\mathrm{C_{13}}$-$\mathrm{C_{15}}$. Such a 2D attractor is also
found for the
interaction $\mathrm{C_{16}}\cdots\mathrm{C_{1}}$-$\mathrm{C_{2}}$
which is 40~$\%$ less intense than the
former and somewhat less than $\mathrm{S_1}\cdots\mathrm{Cl_{2}}$,
$\mathrm{S_1}\cdots\mathrm{Cl_{4}}$, $\mathrm{Cl_{3}}\cdots
\mathrm{C_{4}}$ and $\mathrm{Cl_{1}}\cdots\mathrm{C_{6}}$. These
four latter atom-atom interactions show that $\mathrm{Cl}$ plays
an important role for the electronic coupling along the chains.
The acceptor in the middle (left) of Fig.~\ref{cpintra} is coupled 
to TTF on bottom (top) by $\mathrm{Cl_{1}},\mathrm{Cl_{3}}\cdots
\mathrm{C_{6}},\mathrm{C_{4}}$ ($\mathrm{S_2}$) on the left side
and $\mathrm{Cl_{2}},\mathrm{Cl_{4}}\cdots\mathrm{S_1}$ 
($\mathrm{C_{5}},\mathrm{C_{3}}$) on the right side.

In the low temperature phase (40~K), the pairing of the molecules in
DA pairs is clearly visible on these local contacts (bottom of the
figure on the right part of Fig.~\ref{cpintra} and 
Table~\ref{tab:cpintra}). 
The $\mathrm{S_4}\cdots\mathrm{C_{13}}$-$\mathrm{C_{15}}$ 
contact remains the dominant one with a 50~$\%$ increase 
of its strength. Due to molecular reorientation, the contact
$\mathrm{C_{16}}\cdots\mathrm{C_{1}}$-$\mathrm{C_{2}}$ has changed into 
a more atom-atom like contact between $\mathrm{C_{2}}$ and
$\mathrm{C_{14}}$. Its intensity has more than doubled and with
$\mathrm{S_4}\cdots\mathrm{C_{13}}$-$\mathrm{C_{15}}$ they dominate all 
other local contacts. The latter have also increased by about 35~$\%$
and additionnal strong contacts $\mathrm{O_{2}}\cdots\mathrm{S_{2}}$
and $\mathrm{S_3}\cdots\mathrm{Cl_2}$ have appeared.
These contacts confirm~\cite{lecointe} that inside the pair, 
the CA molecule is essentially connected to the left part of the 
TTF molecule below. Between the DA pairs (top of the right part of 
Fig.~\ref{cpintra}), we observe a small decrease and a more atom-atom
like contact for $\mathrm{S_3}\cdots\mathrm{C_{14}}$-$\mathrm{C_{16}}$
whereas the $\mathrm{S}\cdots\mathrm{Cl}$ and
$\mathrm{Cl}\cdots\mathrm{C}$ contacts moderately increase,
$\mathrm{Cl_4}\cdots\mathrm{C_3}$ being the strongest one.
\begin{table}
\caption{\label{tab:cpintra} Strongest intra-chain contacts for
high (300~K) and low temperature (40~K)
phases. $\epsilon$ and $\mathrm{V(\mathbf{r}_{CP})}$ give
respectively the ellipticity and potential energy density
(kJ. mol$^{-1}$) at each bond critical point. At 40~K, intra and
inter stend for interactions inside a DA pair and between two DA
pairs respectively.}
\begin{ruledtabular}
\begin{tabular}{lcccccc}
 &\multicolumn{2}{c}{300~K}&\multicolumn{2}{c}{40~K intra}
&\multicolumn{2}{c}{40~K inter}\\
&$\epsilon$&$\mathrm{V(\mathbf{r}_{CP})}$&
 $\epsilon$&$\mathrm{V(\mathbf{r}_{CP})}$&
 $\epsilon$&$\mathrm{V(\mathbf{r}_{CP})}$\\
\hline
$\mathrm{S_4}\cdots\mathrm{C_{13}}$-$\mathrm{C_{15}}$& 
7.4& -9.3& 5.8 & -16.5 &     &     \\
$\mathrm{S_3}\cdots\mathrm{C_{14}}$-$\mathrm{C_{16}}$&
7.4& -9.3  &     &      & 4.4 & -7.8\\
\hline
$\mathrm{S_1}\cdots\mathrm{Cl_2}$& 1.3 & -7.0 & 1.6 & -10.6\\
$\mathrm{S_2}\cdots\mathrm{Cl_1}$& 1.3 & -7.0 &&& 1.3 & -7.8\\
\hline
$\mathrm{Cl_3}\cdots\mathrm{C_4}$& 1.3 & -6.7 & 1.3 & -8.9 \\
$\mathrm{Cl_4}\cdots\mathrm{C_3}$& 1.3 & -6.7 &&& 1.2 & -8.7\\
\hline
$\mathrm{O_2}\cdots\mathrm{S_2}$&&& 2.1 & -8.9\\
\hline
$\mathrm{Cl_1}\cdots\mathrm{C_6}$& 2.2 & -6.0 & 2.5 & -8.1 \\
$\mathrm{Cl_2}\cdots\mathrm{C_5}$& 2.2 & -6.0 &&& 1.9 & -6.8\\
\hline
$\mathrm{S_3}\cdots\mathrm{Cl_2}$&&& 2.2 & -8.1\\
\hline
$\mathrm{S_1}\cdots\mathrm{Cl_4}$& 1.3 & -6.0 & 1.6 & -7.8 \\
$\mathrm{S_2}\cdots\mathrm{Cl_3}$& 1.3 & -6.0 &&& 1.1 & -7.1\\
\hline
$\mathrm{C_{16}}\cdots\mathrm{C_{1}}$-$\mathrm{C_{2}}$& 6.0 & -5.6 & \\
$\mathrm{C_{14}}\cdots\mathrm{C_{2}}$&&& 4.5 & -11.7 & \\
\end{tabular}
\end{ruledtabular}
\end{table}

\subsection{\label{sub:inter}Inter-chain interactions}

\begin{figure*}
\resizebox{14.0cm}{!}{
{\includegraphics{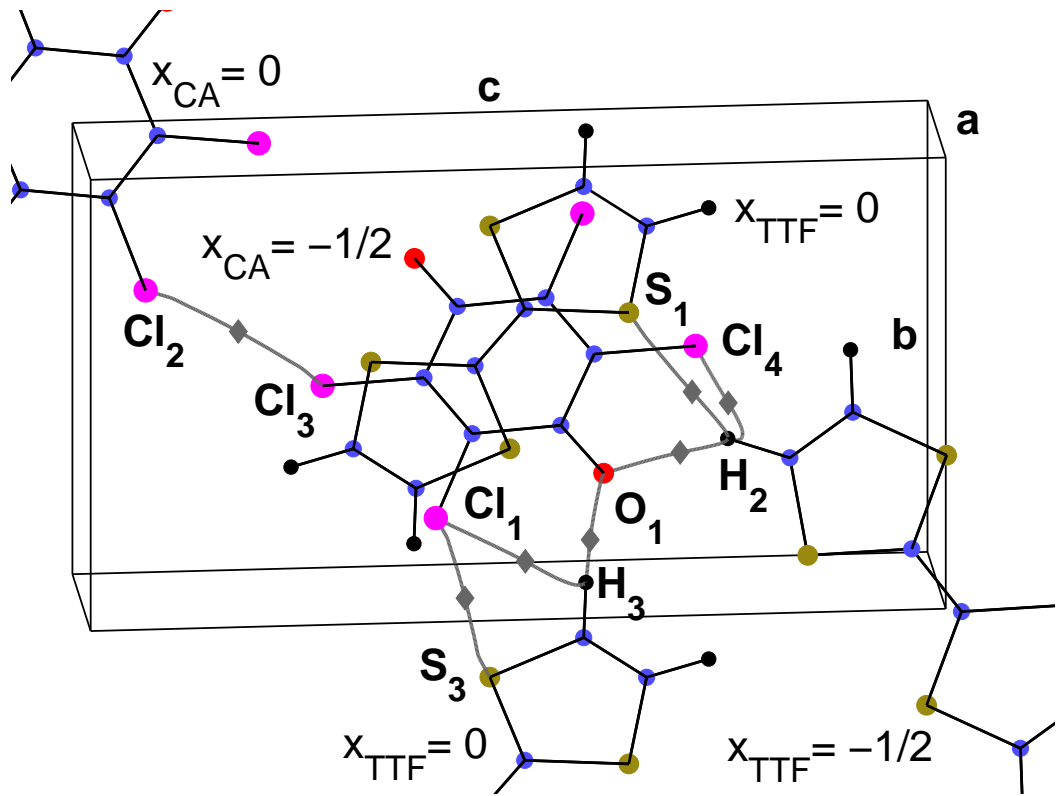}}
}
\caption{\label{cpinter} Inter-chain interactions in the
high temperature phase (300~K).
Diamons correspond to the bond critical points 
and lines indicate the bond path associated to each critical point.
}
\end{figure*}

The inter-chain interactions can be
split in two groups, the first corresponding to interactions between
two chains separated by a primitive translation along $\mathbf{b}$ and 
the second to the interactions between two chains related by the 
gliding plane. In the $\mathbf{c}$ direction the chains are too
far from each other to allow direct atom-atom interactions.
In Fig.~\ref{cpinter} and Table~\ref{tab:cpinter} we have reported the 
properties of the strongest inter-chain contacts. 
For all of them except the two
$\mathrm{Cl}\cdots\mathrm{H}$ contacts, $\epsilon$ is quite
small and the interactions are diatomic. On Fig.~\ref{cpinter}, it
is clear that $\mathrm{Cl_4}$ interacts with the 
$\mathrm{C}-\mathrm{H_2}$ bond. 
In TTF-CA, the importance of hydrogen bonds has already 
been underlined.~\cite{batail,lecointe,oison} Our results show clearly 
that they dominate all other intra and inter-chain interactions
in both high and low temperature phases. $\mathrm{O_1}\cdots\mathrm{H_3}$
in the OH chains is stronger than $\mathrm{O_1}\cdots\mathrm{H_2}$ 
as it is shorter and more straight, the CA and TTF molecular planes 
being aligned in the $\mathbf{b} \pm \mathbf{a}/2$ direction.
The other strong interactions reported in Table~\ref{tab:cpinter}
have comparable $\mathrm{V(\mathbf{r}_{CP})}$'s to the intra-chain ones
given in Table~\ref{tab:cpintra} but there are fewer
of them in the same energy range. 

In the low temperature
phase, the loss of inversion symmetry leads to two inequivalent distances 
for each contact. As already discussed
by Le Cointe et al.~\cite{lecointe}, some of the 
distance reductions (Table~\ref{tab:cpinter}) are larger
than those expected from the thermal unit cell contraction  which is
about 1~$\%$ along $\mathbf{b}$ and $\mathbf{c}$ and 3~$\%$ along 
$\mathbf{a}$ directions. The hydrogen bonds are the most affected and 
become 80~$\%$ and 50~$\%$ stronger for the short and long bonds 
respectively. All other interactions except the long 
$\mathrm{S}\cdots\mathrm{H_2}$ bond, undergo a smaller strengthening. 

\begin{table}
\caption{\label{tab:cpinter} Strongest inter-chain contacts for
high (300~K) and low temperature (40~K)
phases. $\mathrm{V(\mathbf{r}_{CP})}$ corresponds to the
potential energy density (kJ. mol$^{-1}$) at the bond critical point.
d is the distance in \AA\ between the two atoms of
the first column. $\Delta$d gives the relative variation of 
this distance with respect to its value at 300~K.}
\begin{ruledtabular}
\begin{tabular}{lcccccc}
 &\multicolumn{2}{c}{300~K}&\multicolumn{2}{c}{40~K short}
&\multicolumn{2}{c}{40~K long}\\
& $\mathrm{V(\mathbf{r}_{CP})}$ & d
& $\mathrm{V(\mathbf{r}_{CP})}$ & $\Delta$d 
& $\mathrm{V(\mathbf{r}_{CP})}$ & $\Delta$d \\
\hline
$\mathrm{O_1}\cdots\mathrm{H_3}$&-16.3&2.35&-26.7&-7~$\%$&-24.3&-5~$\%$\\
$\mathrm{S_3}\cdots\mathrm{Cl_1}$&-6.8&3.57&-10.3&-5~$\%$& -8.7&-3~$\%$\\
$\mathrm{Cl_1}\cdots\mathrm{H_3}$&-6.7&3.03&-10.8&-6~$\%$& -8.0&-1~$\%$\\
\hline
$\mathrm{O_1}\cdots\mathrm{H_2}$ &-12.0&2.50&-21.4&-8~$\%$&-18.3&-6~$\%$\\
$\mathrm{Cl_4}\cdots\mathrm{C_4}-\mathrm{H_2}$  
& -8.8&3.53&-11.6&-3~$\%$&-11.0&-2~$\%$\\
$\mathrm{S_1}\cdots\mathrm{H_2}$ & -7.8&2.93&-10.4&-4~$\%$& -7.1&+1~$\%$\\
$\mathrm{Cl_2}\cdots\mathrm{Cl_3}$ 
& -7.5&3.53& -9.1&-2~$\%$& -8.1&-1~$\%$\\
\end{tabular}
\end{ruledtabular}
\end{table}

\subsection{Discussion}

Most of the contacts discussed in the Section~\ref{sub:intra} 
and \ref{sub:inter} have already been pointed out on the basis of
a detailed comparison between interactomic distances and van der
Waals radii.~\cite{lecointe} Our results demonstrate that a
topological analysis of the total charge density 
allows to go a step further. It leads to a very precise picture 
of the atoms involved in the different contacts (bond path)
and provides a quantative way to compare their shape ($\epsilon$)
and intensity ($\mathrm{V(\mathbf{r}_{CP})}$). 
From Table~\ref{tab:cpinter} for the inter-chain interactions 
and Table~\ref{tab:cpintra} and Ref.~\onlinecite{lecointe} for 
the intra-chain ones, we find the following relation between
the relative variations of the potential energy density
$\Delta\mathrm{V(\mathbf{r}_{CP})}= \mathrm{(V_{40K} - V_{300K})/
V_{300K}}$ and the length of the bonds 
$\Delta\mathrm{d = (d_{40K} - d_{300K})/d_{300K}}$:
\begin{eqnarray}
\Delta\mathrm{V(\mathbf{r}_{CP})}\simeq  -10\times\Delta\mathrm{d}.
\end{eqnarray}
Thus, the strengthening (diminution) of a given contact defined
by $\mathrm{V(\mathbf{r}_{CP})}$ is directly correlated to the reduction
(augmentation) in distance.

All our results  on TTF-CA show that HOMO-LUMO overlap and hydrogen 
bonds dominate the intermolecular interactions in both high 
and low temperature structures and drive the molecular deformation 
and reorientation occuring at the NIT. The other inter-chain 
contacts do not show any peculiar behavior and are
dragged along with the former.
In addition, many strong interactions inside and between the chains 
involve Cl atoms which must largely contribute to
stabilize these crystal structures.
\section{\label{sec:CT}Charge Transfer}

CT is a key factor of the NIT but it is
is well known that it is not a well defined quantity and 
there are numerous ways to extract it from experimental or theoretical
results. In TTF-CA it has already been estimated from the intensity
of the CT absorption band~\cite{jacobsen}, from the dependence on
molecular ionicity of either bond lengths~\cite{lecointe} 
or vibrational frequencies.~\cite{girlando2}
In this section, we will use three different techniques
to estimate the CT between TTF and CA. The first two are based
on our ab-initio calculations in the crystal while the third is
obtained from a model build on isolated molecules calculations.

\subsection{From the tight-binding model}

Within our tight-binding model,
the CT in the VB is obtained 
by integrating the analytic expression of 
$\vert \mathrm{C_A^{lk}}\vert^2$ (Eq.~(\ref{eqrhovb})).
We have first estimated $\mathrm{\rho_{VB}}$ by using the parameters 
determined in Section~\ref{subsec:tb}, leaving out all those which are 
smaller than 0.005~eV in both phases. A second estimate has been
obtained by keeping only the parameters related to the dispersion
along $\Gamma\rightarrow\mathrm{X}$. In that case, 
$\vert \mathrm{C_A^{lk}}\vert^2$ is given by
\begin{eqnarray}
\mathrm{
|C_A^{lk}|^2} &=& \frac{1}{2} \label{eqcak2}\\
             &-&\mathrm{\frac{(E^A - E^D)}{2\sqrt{(E^A - E^D)^2 +
   4t^2 sin^2\frac{\mathbf{k}\mathbf{a}}{2} +
   4\varepsilon^2 cos^2\frac{\mathbf{k}\mathbf{a}}{2}}}}.\nonumber
\end{eqnarray}
Both approaches lead to the same values of $\mathrm{\rho_{VB}}$
indicated in Table~\ref{tab:CT} where the corresponding experimental
values have also been reported in the last column. Even though 
experiments lead to coarse estimates of the CT as its determination
is always indirect, our high temperature value seems too large.
Two reasons can be put forward to explain this discrepancy:
(i) the tight-binding model is based on a linear combination of
molecular orbitals of isolated molecules
which disregards their deformation due to intermolecular
interactions in the crystal and (ii) the underestimate of the dynamical 
part of the electron-electron interactions due to the LDA approximation.
Among other things, the later will affect the band
gap which is directly included via $\mathrm{E^D - E^A}$ in
Eq.~\ref{eqcak2}. In order to get a feeling for the influence of the
band gap on $\mathrm{\rho_{VB}}$, we increased it by 0.5~eV in both
phases to locate it approximately near the first absorption
peak.\cite{torrance} 
The corresponding values have also been reported in Table~\ref{tab:CT}.
In the high temperature phase, it produces a significant decrease of 
0.16~e$^-$ whereas it is only of 0.03~e$^-$ in the low temperature phase.
Thus $\mathrm{\rho_{VB}}$ as well as its variation between 300 and
40~K becomes in much better agreement with the experimental results.

\begin{table}
\caption{\label{tab:CT} 
Estimates of the charge transfer (CT) from TTF to CA. 
}
\begin{ruledtabular}
\begin{tabular}{lccccc}
&Tight&Tight&Integration&Isolated
&exp.\footnote{From Ref.~\onlinecite{jacobsen}}\\
&Binding&Binding&Atomic&molecules\\
&$\mathrm{E_g}$ & $\mathrm{E_g}+0.5~\mathrm{eV}$&Basin&Model\\
\hline
300~K& 0.59& 0.43& 0.48& 0.47 & 0.2$\pm$0.1\\
40~K& 0.72& 0.69& 0.64& 0.60 & 0.7$\pm$0.1\\
\end{tabular}
\end{ruledtabular}
\end{table}

\subsection{From Bader's theory}

Within Bader's approach, a basin can
be uniquely associated to each atom. It is defined as the region
containing all gradient paths terminating at the atom. The
boundaries of this basin are never crossed by any gradient
vector trajectory and atomic moments are obtained by integration
over the whole basin.\cite{bader,integrity}
The CT between TTF and CA is obtained by summing the atomic charges
belonging to each molecule. For this purpose very high precision
is needed as we are looking for a few tenths of an electron
out of a total of 116 valence electrons. The corresponding CT are
reported in Table~\ref{tab:CT}. These estimates are no more directly 
affected by the band gap error due to the LDA as they are based on the
sole occupied states. Accordingly, they are both at 
300 and 40~K smaller than those obtained by the tight-binding model
and thus closer to the experimental results. Nevertheless, in the
high temperature phase, the CT seems still somewhat overestimated.

\subsection{From a molecular model}

We obtained a third estimate of the CT by means of a simple model
based on isolated molecules calculations which will only be briefly
reported in the present paper. Here, we are looking for the 
charge transfer
$\rho$ which minimizes the total energy in a fixed structure, i.e.
either in the high or low temperature structure. 
The first contribution to the variation of the total energy is the 
molecular one (Carloni et al.~\cite{carloni}):
\begin{eqnarray}
\mathrm{
\Delta E_{mol}(\rho) = \rho(\epsilon_A^0-\epsilon_D^0) 
+ \frac{1}{2} \rho^2 (U_D+U_A)},
\label{eq503}
\end{eqnarray}
where the reference state is the one without any CT, 
$\epsilon_j^0$ is the energy of the relevant orbital in 
the reference state and $\mathrm{U_j}$ the Coulomb repulsion of 
an electron in that orbital. The second contribution 
is the electrostatic one limited to
the Madelung contribution $ q_i q_j / r_{ij}$. The atomic charges
$q_i$ and $q_j$ are 
deduced from PAW calculations on isolated molecules
by using a model density reproducing the multipole moments of
the true molecular charge density.\cite{blo2}
The remaining contributions to the total energy are usually accounted
for by a van der Waals term which is constant in a given structure 
and which can thus be disregarded.
The CT values given by this simple model are also given in 
Table~\ref{tab:CT}. \\

\subsection{Discussion}

It is satisfying that all our three estimates are in close agreement
with each other in both low and high temperature phases. At 40~K,
they are also in good agreement with the experimental value.
At 300~K, we observe a systematic overestimate of about 0.2~e$^-$
with respect to the values obtained by Jacobsen et al.~\cite{jacobsen},
our result being closer to the one deduced from intermolecular
bond lengths (0.4~e$^-$).\cite{lecointe}\\

This discrepancy may have different origins. First of all, the CT 
is not a uniquely defined quantity as its evaluation relies on different 
(and not necessarily equivalent) models based either on theoretical or 
on experimental results. 
Starting from experimental data, the models can only
be very simple and most of the results published in the literature are
based on the behavior of isolated molecules upon ionization: bond-lengths,
intramolecular vibration frequencies, optical spectra..., neglecting thus all
crystalline effects apart from the charge transfer on these molecular
properties.\\

On the other hand, our theoretical results are based on the determination 
of the charge density. They are all affected by the
error due to the LDA approximation to the exchange-correlation
potential. It is well known that this approximation, even in its 
gradient-corrected version, underestimates the dynamical part of 
the electron-electron interactions.  In weakly interacting systems 
where these dynamical effects (or so-called van der Waals interactions) 
are dominant, this leads to an overestimate of the electron delocalization 
and an overbinding effect, together with an underestimate of the band-gap. 
An overestimate of the intermolecular CT is thus expected. However, it 
has already been shown~\cite{oison} that in the ionic phase the multipolar 
static effects are considerably stronger than in the neutral phase, 
so that the LDA error on the CT should also be considerably reduced 
in the low temperature phase.

In our tight-binding model we have clearly evidenced the effect of the 
underestimate of the band gap. 
Within Bader's approach, the CT determination only relies on the 
occupied states, but leads nevertheless to an 
overestimate in the neutral phase. Regarding our model based on isolated 
molecules calculations, it has been shown that the ionization energy 
of TTF~\cite{katanTTF} and the electron affinity of CA~\cite{katanCA}
are correctly obtained within the gradient-corrected LDA approximation 
used in the present study. But our model 
takes into account neither the atomic multipolar contributions
nor the polarization due to the crystal field. 

Thus, our three estimates are coherent and lead probably to a 
correct determination of the CT in the ionic phase, 
and a too large one in the neutral phase. At this stage however, 
the discrepancy with the available experimental values has not 
to be attributed to the sole LDA problem, but also to the definition 
itself of this CT. A good way to check the rather high value obtained 
in this work at 300~K is to perform
an experimental charge density acquisition  by X-ray diffraction 
and use Bader's approach to extract an experimental CT value coherent 
with our theoretical estimate.

\section{Conclusion}

This is the first extensive ab-initio study of the electronic ground
states of TTF-CA, prototype compound of the NIT. We should remind 
here that temperature is included only 
via the experimental
structures, all calculations being performed in a frozen lattice. 

The valence and conduction bands are dominantly of 1D 
character along the stacking 
chains and are essentially a linear combination of the HOMO of
D and the LUMO of A as in TTF-2,5Cl$_2$BQ.\cite{katanjpcm} 
The weak molecular distorsion due to the
symmetry breaking part of the NIT is responsible for the main 
contribution to the charge transfer variation. 
We found also that the thermal lattice contraction leads to a 
decrease of the band gap in the neutral phase, 
going toward a metallic state.
This is consistent with the theoretical work 
of Anusooya-Pati et al.~\cite{anusooya} who found a metallic 
ground state at the transition which is unstable to dimerization. 
It has also to be related with the metallic behavior observed
by Saito et al.~\cite{saito} in an other DA alternating CT
complex.

A detailed analysis of all intermolecular bond critical points
shows that molecular deformation and reorientation occuring at
the NIT are driven by both the HOMO-LUMO overlap and the
hydrogen bonds. Quite many intermolecular contacts involving 
chlorine atoms have also been evidenced. Moreover, we have found
a direct relationship between the relative variations of the
potential energy density at the critical point and the
distance between the atoms involved in the intermolecular
bond. 

Three estimates of the charge transfer have been obtained from
our tight-binding model, from integration over the atomic basins
defined within Bader's approach and from a simple model based
on isolated molecules calculations. All three estimates are in
close agreement with each other and give a value of about 
0.45~e$^-$ at 300~K and 0.65~e$^-$ at 40~K. Experimental charge 
density collection on TTF-CA is under progress in order to
check the rather high value obtained at 300~K.\\

All these results provide a possible picture for the mechanism
of the NIT. The energy dispersions in VB and CB are mainly along 
$\mathbf{a}^\ast$ and $\mathbf{b}^\ast$ with an indirect gap
between $\Gamma$ and $\mathrm{Y}$. Above the transition
temperature, the VB has dominant D character at $\mathrm{Y}$
and the CB a pure A character at $\Gamma$. When temperature is
lowered, the slow decrease of the cell parameters induces a slow 
variation of all transfer integrals and the band gap progressively
vanishes. There is a point where the indirect band gap closes.
Pockets of holes at
$\mathrm{Y}$ and electrons at $\Gamma$ appear which activate a
new mechanism assisted by intra~\cite{katanTTF,venuti}
and intermolecular~\cite{moreac,okimoto} vibrations
creating DA pairs (and longer $\ldots$DADA$\ldots$ strings)
for which the inversion symmetry is locally
broken. This picture is consistent with the observed 
pretransitionnal effects~\cite{mhl} 
described as ionic segments in the neutral 
phase~\cite{okimoto,nagaosa,koshihara,luty} 
sometimes called CT exciton strings.\cite{kuwata}
The rapid increase of CT along the chains may then activate
the hydrogen bonds via intra-molecular
electron-electron interactions between the valence $\pi$
and deep $\sigma$ molecular orbitals.\cite{oison} 
This enables the whole crystal
to topple into the ionic state at the transition
point and is consistent with the sudden decrease of the cell
parameter $\mathbf{b}$.\cite{lecointe} Once the whole crystal has
switched the band gap opens and both VB and CB have strong D-A
character leading to a macroscopic change of the CT.

\begin{acknowledgments}
This work has benefited from collaborations within (1) the
$\Psi_k$-ESF Research Program and (2) the Training and Mobility
of Researchers Program ``Electronic Structure'' (Contract: FMRX-CT98-0178)
of the European Union and (3) the International Joint Research Grant
``Development of charge transfert materials with nanostructures''
(Contract: 00MB4). Parts of the calculations have been supported
by the ``Centre Informatique National de l'Enseignement Sup\'erieur''
(CINES---France). We would like to thank P.E. Bl\"ochl for his PAW
code and A.~Girlando for useful discussions.
\end{acknowledgments}
\bibliography{pr-v7}

\end{document}